# Mass transfer and water management in proton exchange membrane fuel cells with a composite foam-rib flow field

Wei Gao[1], Qifeng Li[1], Kai Sun[1,2], Rui Chen[3], Zhizhao Che[1,2]*, Tianyou Wang[1,2]*

1. State Key Laboratory of Engine, Tianjin University, Tianjin, 300350, China

2. National Industry-Education Platform of Energy Storage, Tianjin University, Tianjin, 300350, China

3. Department of Aeronautical and Automotive Engineering, Loughborough University, Loughborough LE11 3TU, United Kingdom

*Corresponding authors: chezhizhao@tju.edu.cn, wangtianyou@tju.edu.cn

## Abstract

Mass transfer capability of reactants and hydrothermal management is important for the performance and durability of proton exchange membrane fuel cells. In the conventional rib flow field, the oxygen transport is affected by the accumulation of under-rib liquid water which causes excessive concentration loss and limits cell performance. To improve the cell performance, a composite foam-rib flow field structure is proposed by combining the metal foam flow field and the conventional rib flow field. The proposed design is simulated by using a three-dimensional homogeneous non-isothermal numerical model. The results show that the composite foam-rib flow field, by improving the oxygen transfer and water removal capabilities under the ribs, can improve the oxygen concentration and current density without increasing the pumping power, thus improving the cell performance under different conditions. The key parameters of the composite foam-rib flow field are optimized. With the optimal metal foam filling ratio of 0.75 and porosity of 0.85, the peak power density and the limiting current density for the composite foam-rib flow field are higher than the conventional rib flow field by 5.20% and 22.68%.

**Keywords**: Composite foam-rib flow field; Mass transfer; Water management; Proton exchange membrane fuel cell; Fuel cell.

## Nomenclature

| | |
|---|---|
| $a$ | water activity |
| $A$ | cell geometric area ($cm^2$) |
| $C$ | molar concentration ($mol\ m^{-3}$) |
| $C_p$ | specific heat ($J\ kg^{-1}\ K^{-1}$) |
| $D$ | mass diffusivity ($m^2\ s^{-1}$) |
| $EW$ | the equivalent weight of membrane ($kg\ kmol^{-1}$) |
| $F$ | Faraday's constant ($C\ mol^{-1}$) |
| $H$ | height (mm), latent heat ($J\ kg^{-1}$) |
| $I$ | current density ($A\ cm^{-2}$) |
| $j$ | volumetric exchange current density ($A\ m^{-3}$) |
| $\boldsymbol{J}$ | the current density vector ($A\ m^{-2}$) |
| $k$ | thermal conductivity ($W\ m^{-1}\ K^{-1}$) |
| $K$ | permeability ($m^2$) |
| $L$ | length (mm) |
| $\dot{m}$ | mass flow rate ($kg\ s^{-1}$) |
| $M$ | molecular weight ($kg\ mol^{-1}$) |
| $n_d$ | electro-osmotic drag coefficient ($H_2O$ per $H^+$) |
| $P$ | pressure (Pa) |
| $R$ | universal gas constant ($J\ mol^{-1}\ K^{-1}$) |
| $RH$ | relative humidity (%) |
| $S$ | entropy ($J\ mol^{-1}\ K^{-1}$) or source terms |
| $T$ | temperature (K) |
| $\mathbf{u}$ | velocity ($m\ s^{-1}$) |
| $U$ | Oxygen concentration uniformity index |
| $V$ | electrical potential (V) |
| $W$ | width (mm) |
| $X$ | mole fraction |
| $Y$ | mass fraction |

## Greek letters

| | |
|---|---|
| $\varphi$ | volume fraction |
| $\alpha$ | transfer coefficient |
| $\gamma$ | water phase change rate ($s^{-1}$) |
| $\delta$ | thickness (mm) |
| $\varepsilon$ | porosity |
| $\zeta$ | water transfer rate ($s^{-1}$) |
| $\eta$ | overpotential (V) |
| $\theta$ | contact angle (° ) |
| $\kappa$ | electrical conductivity (S $m^{-1}$) |
| $\lambda$ | water content in ionomer |
| $\mu$ | dynamic viscosity (kg $m^{-1}$ $s^{-1}$) |
| $\nu$ | kinetic viscosity ($m^2$ $s^{-1}$) |
| $\xi$ | stoichiometry ratio |
| $\rho$ | density (kg $m^{-3}$) |
| $\sigma$ | surface tension (N $m^{-1}$) |
| $\varphi$ | potential (V) or volume fraction |
| $\omega$ | volume fraction of ionomer in the catalyst layer |

## Subscripts

| | |
|---|---|
| 0 | environmental conditions or intrinsic property |
| a | anode |
| act | activated overpotential |
| a,end | anode bipolar plate |
| bp | bipolar plate |
| c | cathode or capillary pressure |
| c,end | cathode bipolar plate |
| cell | fuel cell |
| CL | catalyst layer |
| cond | condensation |
| d | dissolved water |
| d-g | dissolved water to vapor |
| diff | diffusion |
| d-l | dissolved water to liquid water |
| eff | effective |
| e,ele | electronic |
| equil | equilibrium |
| evap | evaporation |
| EOD | electro-osmotic drag |
| fl | fluid phase |
| g | gas phase |
| GDL | gas diffusion layer |
| g-l | vapor to liquid water |
| $H_2$ | hydrogen |
| i, j | the *i*-th and *j*-th components |
| in | inlet |
| ion | ionic |
| l | liquid water |
| MFR | metal foam rib |
| m | mass or source term |
| mem | membrane |
| mix | mixture |
| out | outlet |
| $O_2$ | oxygen |
| pc | phase change heat (for source term) |
| ref | reference state |
| rev | reversible |
| sat | saturation |
| sl | solid phase |
| surr | surroundings |
| u | momentum (for source term) |
| v | water vapor |
| T | energy (for source term) |

## 1. Introduction

A fuel cell is a high-efficiency energy conversion device that directly converts the chemical energy in hydrogen and oxygen into electrical energy through an electrochemical reaction. Due to its high efficiency and zero emission, it has broad prospects for the sustainable development of green energy [1]. However, before the large-scale commercialization of proton exchange membrane fuel cells (PEMFCs), there are still many challenges in terms of cell performance and durability, one of which is the mass transfer capability of reactants and hydrothermal management [2-6]. To solve this problem and improve cell performance, flow field optimization is considered to be an effective strategy.

Flow field optimization is an important area of research in PEMFCs. Through flow field optimization, the concentration of reactants at the reaction site can be increased, thereby reducing the concentration loss, making the reaction gas distribution more uniform, avoiding the occurrence of "starvation", and at the same time ensuring rapid removal of the excess liquid water from the flow field without dehydration in the polymer. The widely used rib flow field structures are parallel, serpentine, and interdigitated flow fields [7-9]. Each type of rib flow field structure has its advantages and disadvantages. For example, in the parallel flow field, the pressure drop is small, but the flow velocity is uneven, and the performance is unstable. In contrast, the serpentine flow field has uniform flow velocity distribution, and the heat/mass transfer is good, but the pressure drop is too large, and the mass transport in the downstream area of the channels is not sufficient [10]. The interdigitated flow field can enhance the convection of gas and the water removal capability under the ribs, and improve the mass transport and water removal effects; but it also induces a large pressure drop, resulting in more pumping power [11]. These conventional rib flow fields (CRFF) structures can easily cause an accumulation of liquid water discharged from the porous media under the ribs, and at the same time, liquid films are often formed in the corners of the channel with low gas velocity, which further reduces the oxygen transport capability and cell performance. To avoid these problems in conventional rib flow fields, different flow field structures have been proposed by optimizing the number, size, and shape of conventional flow channels [12-17], rearranging flow channels [18-20], redesigning intake channels [21-25], adding baffles in flow channels [26-33], designing bionic flow fields [34-36], and proposing three-dimensional (3D) spatial structure flow field designs [37-39]. For example, Bagherighajari et al. [33] improved cell performance by adding blocks in the gas flow channel on the cathode side to form the intermediate blocked interdigitated flow field structure. By numerical simulation, the results show that the flow field structure can enhance the transfer rate of reactants into the gas diffusion layer, and thus improve the cell performance. These CRFF structures can, to some extent, strengthen the mass transfer and water removal capabilities under the ribs, reduce pumping power, and further improve fuel cell performance.

Compared with conventional rib flow fields, 3D porous flow field structure, such as metal foam [40], metal fiber [41], carbon foam [42], carbon felt [43], and graphene foam [44], is more conducive to oxygen transport and uniform distribution of reactants [45]. Azarafza et al. [46] compared the parallel flow fields, serpentine flow fields, interdigitated flow fields, baffle parallel flow fields, and metal foam flow fields, through numerical simulation, and the results showed that the current density and oxygen concentration distribution for the metal foam flow field is relatively uniform, the pressure drop is relatively moderate, and it has better cell performance. Jo et al. [47] compared the water management performance of the metal foam flow field and serpentine flow field by numerical simulation, and the results showed that the metal foam flow field has higher cell performance compared with the serpentine flow field, but is more suitable for working under low humidity operating condition. This is because the convection in the metal foam in the metal foam flow field is weak and the effect on water removal in the downstream part of the channels is not obvious. Lee et al. [48] found that under the condition of excess dry air supply, the metal foam flow field has a better water-retention capability and more uniform current density and reactant distribution than the conventional rib flow field. Zhang et al. [49] proposed convergent-inlet divergent-outlet manifold channels to alleviate the low-speed problem in the corner area of the metal foam flow field and achieved more uniform flow distribution and lower pressure drop. To sum up, although the metal foam flow field has better cell performance than the conventional rib flow field, it is easy to form a weak flow zone due to its lack of mainstream direction, resulting in a waste of reaction area. In addition, liquid water removal is difficult due to the weak convection induced in the flow field by the complex 3D porous structure.

To solve those problems in metal foam flow fields, many designs were proposed. Kang et al. [50] used metal foams with different pore gradients to optimize the metal foam flow field through local non-uniform design. The results showed that the peak power density of the metal foam flow field with an appropriate pore gradient is 8.23% higher than that without a pore gradient. Deng et al. [51] proposed a porous-rib flow field structure by grooving flow channels in the metal foam. Compared with the conventional rib flow field structure, the proposed porous-rib flow field can reduce the oxygen transport resistance under the ribs. Hence it can increase the peak power density by 9%, increase the limit current density by 15%, and reduce the pressure drop by 38%. Bilondi et al. [52] studied the effect of porous carbon inserts (PCI) structure on the cell performance by CFD simulations. The results showed that the optimal value of cell performance exists when the PCI content is 80%, which can improve cell performance by 23%. In addition, over the study range, cell performance is improved by increasing the PCI porosity and the rib/channel width ratios. To guide the flow in metal foam, Kariya et al. [53] proposed to install different separators in the porous flow field to improve the uniform oxygen supply capability. The power density of the separator installed in the porous flow field is higher, but its pressure loss is larger than that without separators. Son et al. [54] designed a serpentine structure on the metal foam flow field of the cathode to guide the reactants to the corners of the reaction zone. Compared with the metal foam flow field, the cell performance was improved by 4.7% under the condition of 2 turns serpentine structure at 0.5 V. With the increase of the number of turns of the serpentine structure, the concentration of the reactant at the reaction site increases, but the pumping power is also higher.

Therefore, in the design of flow field structure in PEMFCs, it is necessary to further enhance the capabilities of oxygen transport and water removal under the ribs, improving the cell performance without increasing the pumping power. Moreover, the guiding fluid effect of the ribs and the convective transfer effect under the ribs are important for the distribution of the reactants. In this study, a composite foam-rib flow field (CFRFF) structure is proposed by replacing part of the rib of the CRFF with metal foam. Our analysis shows that the design can enhance the oxygen transfer and water removal capabilities under the ribs, further improving cell performance without increasing the pumping power. A 3D homogeneous non-isothermal numerical model is adopted to study the flow distribution in the CFRFF, and the cell performance under different operating conditions and structure parameters are analyzed. The mechanism for the CFRFF structure to enhance the cell performance is unveiled, and the structural parameters of the CFRFF are optimized.

## 2. Geometries and numerical simulation method

### 2.1 Model geometries

The CFRFF is proposed by replacing part of the rib of the conventional rib flow field with metal foam. It consists of bipolar plates (BPs), gas flow channels (GFCs), metal foam ribs (MFRs), gas diffusion layers (GDLs), microporous layers (MPLs), catalytic layers (CLs), and a proton exchange membrane (PEM). The cathode is the CFRFF structure, and 50% of the rib is filled with nickel metal foam with a porosity of 95% (unless explicitly specified otherwise), as shown in Fig. 1a. The anode is the conventional parallel straight channel flow field, as the transfer of hydrogen is faster and usually there is no water-flooding issue on the anode side. For comparison, a PEMFC model with the CRFF is also built, as shown in Fig. 1b. All parameters for the two models are the same unless otherwise explicitly stated, and they are listed in Table 1. The origin of 3D geometric coordinates is located at the entrance of the center line of the cathode end face along the flow direction. The single-channel geometric model is used, and a homogeneous model is adopted to treat the porous media.

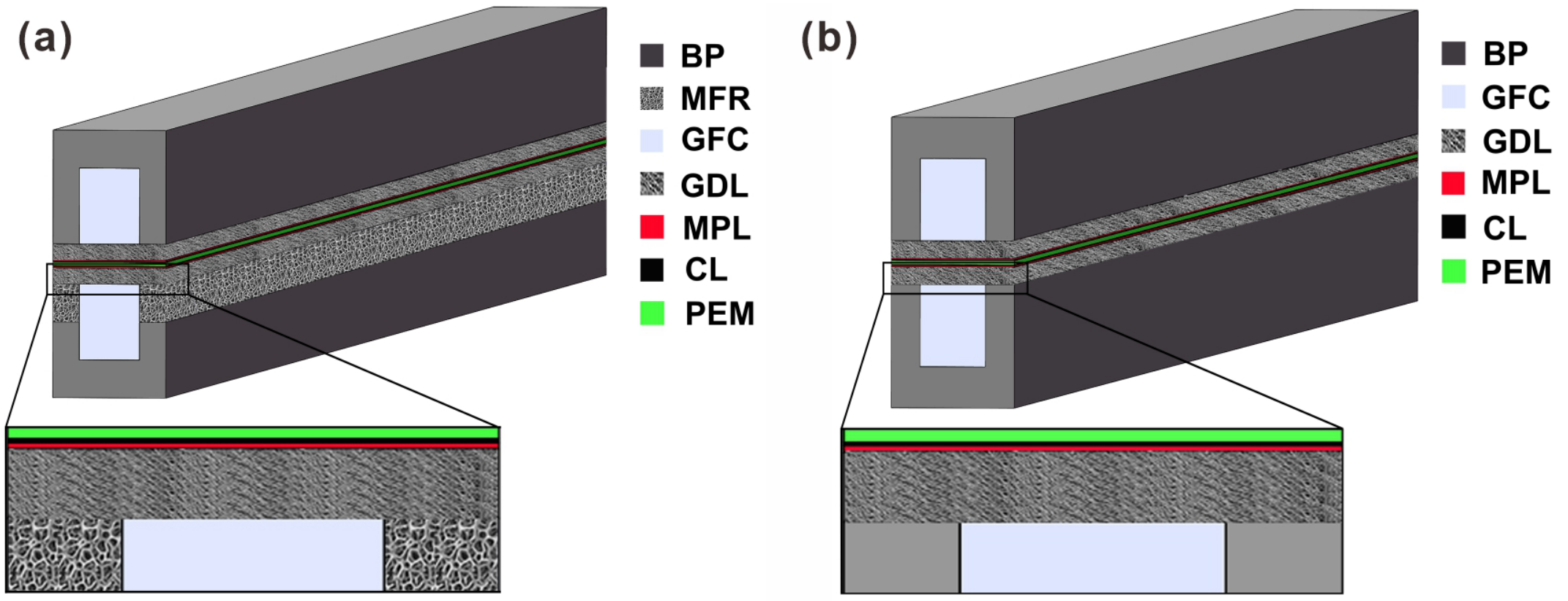

Fig. 1. (a) CFRFF and (b) CRFF structure models used for the numerical simulation.

Table 1 Parameters of the PEMFC model considered in this study.

| Parameters | Symbol | Values |
| --- | --- | --- |
| BP length, width, height (mm) [55] | $L_{bp}$, $W_{bp}$, $H_{bp}$ | 50, 1.5, 0.5 |
| Rib length, width, height (mm) [55] | $L_{rib}$, $W_{rib}$, $H_{rib}$ | 50, 0.7, 1.0 |
| GFC length, width, height (mm) [55] | $L_{ch}$, $W_{ch}$, $H_{ch}$ | 50, 0.8, 1.0 |
| GDL thickness (μm) [55] | $\delta_{GDL}$ | 210 |
| MPL thickness (μm) [55] | $\delta_{MPL}$ | 20 |
| CL thickness (μm) [55] | $\delta_{CL}$ | 10 |
| PEM thickness (μm) [55] | $\delta_{mem}$ | 50.8 |
| MFR height (mm) | $H_{MFR}$ | 0.5 |
| Typical pore diameter of metal foam (mm) | $d_{MFR}$ | 0.2 |
| Equivalent weight of PEM (kg kmol$^{-1}$) [55] | $EW$ | 1100 |
| Ionomer volume fraction in CL [55] | $\omega$ | 0.27 |
| Ionomer volume fraction in PEM [55] | $\omega$ | 1.0 |
| Surface tension coefficient (N m$^{-1}$) [55] | $\sigma$ | 0.0625 |
| Contact angle (MFR, GDL, MPL, CL) (°) [47, 55] | $\theta$ | 120, 120, 115, 100 |
| Porosity (MFR, GDL, MPL, CL) [49, 56] | $\varepsilon$ | 0.95, 0.6, 0.5, 0.3 |
| Intrinsic permeability (MFR, GDL, MPL, CL) (m$^2$) [49, 55] | $K_0$ | 1e-08, 1e-11, 1e-12, 1e-13 |
| Electrical conductivity (BP, MFR, GDL, MPL, CL) (S m$^{-1}$) [49, 55] | $\kappa_e$ | 20000, 1.4e07, 5000, 2000, 2000 |
| Thermal conductivity (BP, MFR, MPL, CL, PEM) (W m$^{-1}$ K$^{-1}$) [49, 55] | $\kappa$ | 20, 91.74, 1, 1, 0.95 |
| Anisotropic thermal conductivity (GDL) (W m$^{-1}$ K$^{-1}$) [57] | $\begin{cases}\kappa_{\text{in-plane}} \\ \kappa_{\text{through-plane}}\end{cases}$ | $\kappa_{\text{in-plane}} = 21,\ \kappa_{\text{through-plane}} = 1.7$ |
| Specific heat capacity(BP, MFR, GDL, MPL, CL, PEM) (J kg$^{-1}$ K$^{-1}$) [49, 55] | $C_p$ | 1580, 440, 568, 3300, 3300, 833 |

### 2.2 Model simplifications and assumptions

The multiphysics process in the fuel cell involves fluid flow, heat transfer, mass transfer, chemical reaction, electron conduction, etc. To facilitate the establishment of a tractable model for such a complex multiphysics process, the PEMFC is simplified based on several assumptions. a) The PEMFC system is operated under stable conditions, and the effect of gravity on the transport is ignored. b) The MFRs, GDLs, MPLs, CLs, and PEM are assumed to be homogeneous, and the effect of porous media anisotropy on the mass transport process is ignored. c) The reaction gases follow the ideal gas law, and the flow is laminar. d) The PEM is not permeable, and the cathode and anode reaction gases cannot be transported across the PEM. e) The liquid water in the gas flow channel is in a mist state, and the liquid water is calculated only in the porous media. f) The gas diffusion in the flow field follows Fick's law, and the Knudsen diffusion in the catalyst layer is not considered.

### 2.3 3D model governing equations

With the assumptions in Section 2.2, the transport process in the PEMFC can be modeled by the mass conservation equation, the momentum conservation equation, the species transport equation, the liquid water saturation equation, the dissolved water transport equation, the electron and ion charge transport equations, and energy conservation equation. The purpose of this study is not to optimize the mathematical model but to focus on the cell performance for the CFRFF and CRFF structure. Therefore, we use the existing two-phase model that has been validated in the literature.

The mass conservation equation in the MFRs, GDLs, MPLs, CLs, and GFCs:

$$\nabla \cdot \left( \rho_{\mathrm{g}} \mathbf{u}_{\mathrm{g}} \right) = S_{\mathrm{m}} \tag{1}$$

The momentum conservation equation in the MFRs, GDLs, MPLs, CLs, and GFCs:

$$\nabla\cdot\left(\frac{\rho_{\mathrm{g}}\mathbf{u}_{\mathrm{g}}\mathbf{u}_{\mathrm{g}}}{\varepsilon^{2}\left(1-\varphi_{\mathrm{l}}\right)^{2}}\right)=-\nabla P_{\mathrm{g}}+\mu_{\mathrm{g}}\nabla\cdot\left[\nabla\left(\frac{\mathbf{u}_{\mathrm{g}}}{\varepsilon\left(1-\varphi_{\mathrm{l}}\right)}\right)+\nabla\left(\frac{\mathbf{u}_{\mathrm{g}}}{\varepsilon\left(1-\varphi_{\mathrm{l}}\right)}\right)^{\mathrm{T}}\right]-\frac{2}{3}\mu_{\mathrm{g}}\nabla\left(\nabla\cdot\left(\frac{\mathbf{u}_{\mathrm{g}}}{\varepsilon\left(1-\varphi_{\mathrm{l}}\right)}\right)\right)+S_{\mathrm{u}} \quad (2)$$

The species transport equation in the MFRs, GDLs, MPLs, CLs, and GFCs:

$$\nabla\cdot\left(\rho_{\mathrm{g}}\mathbf{u}_{\mathrm{g}}Y_{\mathrm{i}}\right)=\nabla\cdot\left(\rho_{\mathrm{g}}D_{\mathrm{i}}^{\mathrm{eff}}\nabla Y_{\mathrm{i}}\right)+S_{\mathrm{i}} \quad (3)$$

The liquid water saturation in porous media (i.e., MFRs, GDLs, MPLs, and CLs) follows:

$$\nabla\cdot\left(\frac{\rho_{\mathrm{l}}K_{\mathrm{l}}}{\mu_{\mathrm{l}}}\nabla(p_{\mathrm{g}}-p_{\mathrm{c}})\right)+S_{\mathrm{l}}=0 \quad (4)$$

The dissolved water transport equation in the PEM and CLs is:

$$\nabla\cdot\left(n_{\mathrm{d}}\frac{\mathbf{J}_{\mathrm{ion}}}{F}\right)=\frac{\rho_{\mathrm{mem}}}{EW}\nabla\cdot\left(D_{\mathrm{d}}^{\mathrm{eff}}\nabla\lambda_{\mathrm{d}}\right)+S_{\mathrm{d}} \quad (5)$$

The electron and ion charge transport equations in BPs, MFRs, GDLs, MPLs, CLs, and PEM are:

$$\nabla\cdot\left(\kappa_{\mathrm{e}}^{\mathrm{eff}}\nabla\varphi_{\mathrm{e}}\right)+S_{\mathrm{e}}=0 \quad (6)$$

$$\nabla\cdot\left(\kappa_{\mathrm{ion}}^{\mathrm{eff}}\nabla\varphi_{\mathrm{ion}}\right)+S_{\mathrm{ion}}=0 \quad (7)$$

The energy conservation equation for BPs, MFRs, GFCs, GDLs, MPLs, CLs, and PEM:

$$\nabla\cdot\left(\left(\rho C_{\mathrm{p}}\right)_{\mathrm{fl}}^{\mathrm{eff}}\mathbf{u}T\right)=\nabla\cdot\left(k_{\mathrm{fl,sl}}^{\mathrm{eff}}\nabla T\right)+S_{\mathrm{T}} \quad (8)$$

The expressions of the source terms involved in the above equations are summarized in Table 2. The expressions involving electrochemical reactions, charge transport, and phase transition processes are shown in Table 3.

### 2.4 Boundary conditions

The boundary conditions at the inlet and the outlet of the anode and cathode flow channels are set as the mass flow inlet and the pressure outlet. The mass flow inlet conditions of the anode and cathode are, respectively:

Table 2 The source terms of equations [5, 55, 56, 58, 59].

| | $S_m$ | $S_u$ | $S_i$ | $S_l$ | $S_e$ | $S_{ion}$ | $S_d$ | $S_T$ | $S_{H2O}$ |
|---|---|---|---|---|---|---|---|---|---|
| BP | 0 | 0 | 0 | 0 | 0 | 0 | 0 | $\left\|\nabla\varphi_c\right\|^2\kappa_e^{eff}$ | 0 |
| GFC | 0 | 0 | 0 | 0 | 0 | 0 | 0 | 0 | 0 |
| MFR | $S_{H_2O}$ | $-\left(\frac{\mu_g}{K_g}\mathbf{u}_g+\beta\rho\left\|\mathbf{u}_g\right\|\mathbf{u}_g+\frac{S_m}{\varepsilon^2}\mathbf{u}_g\right)$ | 0 | $S_{g\text{-}l}$ | 0 | 0 | 0 | $\left\|\nabla\varphi_c\right\|^2\kappa_e^{eff}+S_{pc}$ | $-S_{g\text{-}l}$ |
| GDL | $S_{H_2O}$ | $-\left(\frac{\mu_g}{K_g}\mathbf{u}_g+\beta\rho\left\|\mathbf{u}_g\right\|\mathbf{u}_g+\frac{S_m}{\varepsilon^2}\mathbf{u}_g\right)$ | 0 | $S_{g\text{-}l}$ | 0 | 0 | 0 | $\left\|\nabla\varphi_c\right\|^2\kappa_e^{eff}+S_{pc}$ | $-S_{g\text{-}l}$ |
| MPL | $S_{H_2O}$ | $-\left(\frac{\mu_g}{K_g}\mathbf{u}_g+\beta\rho\left\|\mathbf{u}_g\right\|\mathbf{u}_g+\frac{S_m}{\varepsilon^2}\mathbf{u}_g\right)$ | 0 | $S_{g\text{-}l}$ | 0 | 0 | 0 | $\left\|\nabla\varphi_c\right\|^2\kappa_e^{eff}+S_{pc}$ | $-S_{g\text{-}l}$ |
| ACL | $S_{H_2}+S_{H_2O}$ | $-\left(\frac{\mu_g}{K_g}\mathbf{u}_g+\beta\rho\left\|\mathbf{u}_g\right\|\mathbf{u}_g+\frac{S_m}{\varepsilon^2}\mathbf{u}_g\right)$ | $S_{H_2}=-\frac{j_a}{2F}M_{H_2}$ | $S_{g\text{-}l}$ | $-j_a$ | $j_a$ | $-S_{d\text{-}v}-\frac{\rho_l K_{mem}(\bar{p}_{la}-\bar{p}_{lc})}{\mu_l M_{H_2O}\delta_{mem}\delta_{CL}}$ | $j_a\left\|\eta_{act}\right\|+\frac{j_a T\Delta S}{2F}+\left\|\nabla\varphi_e\right\|^2\kappa_e^{eff}+\left\|\nabla\varphi_{ion}\right\|^2\kappa_{ion}^{eff}+S_{pc}$ | $-S_{g\text{-}l}+S_{d\text{-}v}M_{H_2O}$ |
| CCL | $S_{O_2}+S_{H_2O}$ | $-\left(\frac{\mu_g}{K_g}\mathbf{u}_g+\beta\rho\left\|\mathbf{u}_g\right\|\mathbf{u}_g+\frac{S_m}{\varepsilon^2}\mathbf{u}_g\right)$ | $S_{O_2}=-\frac{j_c}{4F}M_{O_2}$ | $S_{g\text{-}l}+S_{d\text{-}l}$ | $j_c$ | $-j_c$ | $\frac{j_c}{2F}-S_{d\text{-}l}+\frac{\rho_l K_{mem}(\bar{p}_{la}-\bar{p}_{lc})}{\mu_l M_{H_2O}\delta_{mem}\delta_{CL}}$ | $j_c\left\|\eta_{act}\right\|+\frac{j_c T\Delta S}{4F}+\left\|\nabla\varphi_e\right\|^2\kappa_e^{eff}+\left\|\nabla\varphi_{ion}\right\|^2\kappa_{ion}^{eff}+S_{pc}$ | $-S_{g\text{-}l}$ |
| PEM | 0 | 0 | 0 | 0 | 0 | 0 | 0 | $\left\|\nabla\varphi_{ion}\right\|^2\kappa_{ion}^{eff}$ | 0 |

Table 3 Correlations related to electrochemical reaction, proton, and ion transport, and phase change processes.

| Parameters | Correlations |
|---|---|
| Effective mass diffusion coefficient ($m^2 s^{-1}$) [55] | $D_i^{eff} = \varepsilon^{1.5}(1-\varphi_l)^{1.5} D_i$ |
| The ideal mixed gas density (kg $m^{-3}$) [55] | $\rho_g = p_g \left( RT \sum_i \frac{Y_i}{M_i} \right)^{-1}$ |
| The gas relative permeability ($m^2$) [55] | $K_g = K_0 (1-\varphi_l)^{3.0}$ |
| The liquid relative permeability ($m^2$) [55] | $K_l = K_0 \varphi_l^{3.0}$ |
| The capillary pressure (Pa) [55, 60] | $P_c = \begin{cases} \sigma \cos\theta \left( \frac{\varepsilon}{K_0} \right)^{0.5} \left[ 1.417(1-\varphi_l) - 2.12(1-\varphi_l)^2 + 1.26(1-\varphi_l)^3 \right], \theta_c < 90^\circ \\ \sigma \cos\theta \left( \frac{\varepsilon}{K_0} \right)^{0.5} \left( 1.417\varphi_l - 2.12\varphi_l^2 + 1.26\varphi_l^3 \right), \theta_c > 90^\circ \end{cases}$ |
| The ion current density calculation vector (A $m^{-2}$) [55] | $\mathbf{J}_{ion} = -\kappa_{ion}^{eff} \nabla \phi_{ion}$ |
| Anode active overpotential (V) [55] | $\eta_{a,act} = \varphi_e - \varphi_{ion}$ |
| Cathodic active overpotential (V) [55] | $\eta_{c,act} = \varphi_e - \varphi_{ion} - V_{rev}$ |
| The reversible voltage (V) [55] | $V_{rev} = 1.229 - 0.9 \times 10^{-3} (T - 298.15) + \frac{RT}{2F} \left( \ln P_{H_2} + \frac{1}{2} \ln P_{O_2} \right)$ |
| Ionic conductivity (S $m^{-1}$) [61] | $\kappa_{ion} = \begin{cases} (0.3300\lambda_d + 0.010) \exp\left[ 1268 \left( \frac{1}{303.15} - \frac{1}{T} \right) \right], \lambda_d \le 1 \\ (0.5139\lambda_d - 0.326) \exp\left[ 1268 \left( \frac{1}{303.15} - \frac{1}{T} \right) \right], \lambda_d > 1 \end{cases}$ |
| Effective ionic conductivity (S $m^{-1}$) [62, 63] | $\kappa_{ion}^{eff} = \omega^{1.5} \kappa_{ion}$ |
| Effective electronic conductivity (S $m^{-1}$) [62, 63] | $\kappa_{ele}^{eff} = (1 - \varepsilon - \omega)^{1.5} \kappa_{ele}$ |
| Anode/Cathode transfer coefficient [55] | $\alpha_a = \alpha_c = 0.5$ |
| Reference exchange current density (A $m^{-3}$) [55] | $j_{0,a}^{ref} = 10^8 \exp\left[ -1400 \left( \frac{1}{T} - \frac{1}{353.15} \right) \right]$ |
| Reference exchange current density (A $m^{-3}$) [55] | $j_{0,c}^{ref} = 120 \exp\left[ -7900 \left( \frac{1}{T} - \frac{1}{353.15} \right) \right]$ |
| Hydrogen reference concentration (mol $m^{-3}$) [55] | $C_{H_2}^{ref} = 56.4$ |
| Oxygen reference concentration (mol $m^{-3}$) [55] | $C_{O_2}^{ref} = 3.39$ |
| Electro-osmotic drag coefficient [55] | $n_d = 2.5\lambda_d / 22$ |
| The Butler-Volmer equation (A $m^{-3}$) [55] | $\begin{cases} j_a = (1-\varphi_l) j_{0,a}^{ref} \left( \frac{(1-\varphi_l)\varepsilon C_{H_2}}{C_{H_2}^{ref}} \right)^{0.5} \left( \exp\left( \frac{2F\alpha_a}{RT}\eta_{act} \right) - \exp\left( -\frac{2F\alpha_c}{RT}\eta_{act} \right) \right) \\ j_c = (1-\varphi_l) j_{0,c}^{ref} \left( \frac{(1-\varphi_l)\varepsilon C_{O_2}}{C_{O_2}^{ref}} \right)^{3} \left( -\exp\left( \frac{4F\alpha_a}{RT}\eta_{act} \right) + \exp\left( -\frac{4F\alpha_c}{RT}\eta_{act} \right) \right) \end{cases}$ |
| Liquid water and water vapor phase change source (kg $m^{-3}$ $s^{-1}$) [64] | $S_{g\text{-}l} = \begin{cases} \zeta_{cond}\varepsilon(1-\varphi_l) M_{H_2O} \frac{(P_v - P_{sat})}{RT}, P_v > P_{sat} \\ \zeta_{evap}\varepsilon\varphi_l M_{H_2O} \frac{(P_v - P_{sat})}{RT}, P_v < P_{sat} \end{cases}$ |
| Absorption of water vapor and membrane water (mol $m^{-3}$ $s^{-1}$) [65] | $S_{d\text{-}v} = \frac{k_{ab}}{\delta_{CL}} \frac{\rho_{mem}}{EW} (\lambda_d - \lambda_{equil})$ |
| Desorption of liquid and dissolved water (mol $m^{-3}$ $s^{-1}$) [65] | $S_{d\text{-}l} = \frac{k_{de}}{\delta_{CL}} \frac{\rho_{mem}}{EW} (\lambda_d - \lambda_{equil})$ |
| Absorption coefficient of water vapor and membrane water (m $s^{-1}$) [65] | $k_{ab} = 1.14\times10^{-5} \frac{M_{H_2O}\lambda_d/\rho_l}{EW/\rho_{mem} + M_{H_2O}\lambda_d/\rho_l} \exp[2416(\frac{1}{303} - \frac{1}{T})]$ |
| Desorption coefficient of liquid and dissolved water (m $s^{-1}$) [65] | $k_{de} = 4.59\times10^{-5} \frac{M_{H_2O}\lambda_d/\rho_l}{EW/\rho_{mem} + M_{H_2O}\lambda_d/\rho_l} \exp[2416(\frac{1}{303} - \frac{1}{T})]$ |
| Saturation vapor pressure (Pa) [56] | $\log_{10}(P^{sat}/101325) = -2.1794 + 0.02953(T - 273.15) - 9.183\times10^{-5}(T-273.15)^2 + 1.4454\times10^{-7}(T-273.15)^3$ |
| Diffusivity of dissolved water ($m^2$ $s^{-1}$) [55] | $D_d = \begin{cases} 3.10\times10^{-7}\lambda_d[\exp(0.28\lambda_d) - 1]\exp\left(\frac{-2346}{T}\right), 0 < \lambda_d < 3 \\ 4.17\times10^{-8}\lambda_d[161\exp(-\lambda_d) + 1]\exp\left(\frac{-2346}{T}\right), 3 \le \lambda_d < 17 \\ 4.10\times10^{-10}(\frac{\lambda_d}{25})^{0.15}(1 + \tanh(\frac{\lambda_d - 2.5}{1.4})), \lambda_d \ge 17 \end{cases}$ |
| Equilibrium dissolved water [56] | $\lambda_{equil} = \begin{cases} 0.043 + 17.81a - 39.85a^2 + 36.0a^3, 0 \le a \le 1 \\ 14.0 + 14(a-1), 1 < a \le 3 \\ 16.8, a > 3 \end{cases}$ |
| Water activity [55] | $a = \frac{X_{vp}P_g}{P_{sat}} + 2\varphi_l$ |
| Oxygen concentration uniformity index [51] | $U_{O_2} = 1 - \frac{1}{C_{O_2,ave}}\sqrt{\frac{1}{A_{ave}}\iint (C_{O_2} - C_{O_2,ave})^2 dA}$ |
| Temperature uniformity index [51] | $U_T = 1 - \frac{1}{T_{ave}}\sqrt{\frac{1}{A_{ave}}\iint (T - T_{ave})^2 dA}$ |
| The metal foam filling ratio ($R$) | $R = H_{MFR}/H_{rib}$ |

$$\dot{m}_{\mathrm{in,a}} = \frac{\rho_{\mathrm{g}}^{\mathrm{a}} \xi^{\mathrm{a}} I_{\mathrm{ref}} A_{\mathrm{mem}}}{2FC_{\mathrm{H_2}}} \tag{9}$$

$$\dot{m}_{\mathrm{in,c}} = \frac{\rho_{\mathrm{g}}^{c} \xi^{\mathrm{c}} I_{\mathrm{ref}} A_{\mathrm{mem}}}{4FC_{\mathrm{O_2}}} \tag{10}$$

The concentrations of oxygen and hydrogen at the inlet of the anode and cathode are, respectively:

$$C_{\mathrm{H_2}} = \frac{\left(P_{\mathrm{g}}^{\mathrm{a}} - RH_{\mathrm{a}} P^{\mathrm{sat}}\right)}{RT_0} \tag{11}$$

$$C_{\mathrm{O_2}} = \frac{0.21\left(P_{\mathrm{g}}^{\mathrm{c}} - RH_{\mathrm{c}} P^{\mathrm{sat}}\right)}{RT_0} \tag{12}$$

The boundary conditions of the bipolar plate potential can be expressed as:

$$\begin{cases} \varphi_{\mathrm{e}}^{\mathrm{a,end}} = 0 \\ \varphi_{\mathrm{e}}^{\mathrm{c,end}} = V_{\mathrm{cell}} \end{cases} \tag{13}$$

The detailed boundary conditions are shown in Table 4. The dynamic viscosity and mass diffusivity of the reaction gas depends on pressure and temperature, and their relationships are provided in Table 5. The main working conditions and parameters of the PEMFCs in this study are detailed in Table 6.

Table 4 List of the imposed boundary conditions.

| | Flow conditions | Species | Liquid water | Charge | Temperature | Other variables |
|---|---|---|---|---|---|---|
| Anode inlet | $\dot{m}_{\mathrm{in,a}} = \frac{\rho_{\mathrm{g}}^{\mathrm{a}} \xi^{\mathrm{a}} I_{\mathrm{ref}} A_{\mathrm{mem}}}{2FC_{\mathrm{H_2}}}$ | $Y_{\mathrm{i}} = Y_{\mathrm{i,in}}$ <br> $i = H_2, H_2O$ | - | - | $T_{\mathrm{in,a}}$ | - |
| Cathode inlet | $\dot{m}_{\mathrm{in,c}} = \frac{\rho_{\mathrm{g}}^{c} \xi^{\mathrm{c}} I_{\mathrm{ref}} A_{\mathrm{mem}}}{4FC_{\mathrm{O_2}}}$ | $Y_{\mathrm{i}} = Y_{\mathrm{i,in}}$ <br> $i = O_2, H_2O, N_2$ | - | - | $T_{\mathrm{in,c}}$ | - |
| Outlets | $p = p_{\mathrm{out}}$ | $-\mathbf{n} \cdot \rho D_{\mathrm{eff,i}} \nabla Y_{\mathrm{i}} = 0$ | - | - | $-\mathbf{n} \cdot \mathbf{q} = 0$ | - |
| Wall | $\mathbf{u} = \mathbf{0}$ | $-\mathbf{n} \cdot \mathbf{J}_{\mathrm{j}} = 0$ | - | $-\mathbf{n} \cdot \mathbf{i} = 0$ | $-\mathbf{n} \cdot \mathbf{q} = 0$ | - |
| Anode terminal | - | - | - | $V_{\mathrm{cell}} = 0\ V$ | $T = 353.15\ K$ | - |
| Cathode terminal | - | - | - | $V_{\mathrm{cell}}$ | $T = 353.15\ K$ | - |
| GDL/GFC interface | - | - | $\varphi_0 = p_{\mathrm{c}}(J(\varphi_{\mathrm{l}}))^{-1}$ / <br> $f_{\mathrm{liq}} = \theta \varepsilon \varphi_{\mathrm{l}} \cdot \max\left[\left(p_{\mathrm{c}} + \frac{1}{2}\rho_{\mathrm{l}} V^2\right), 0\right]$ | - | - | - |
| Symmetry | $\mathbf{u} \cdot \mathbf{n} = 0$ | $-\mathbf{n} \cdot \mathbf{J}_{\mathrm{j}} = 0$ | $-\mathbf{n} \cdot \mathbf{N}_{\mathrm{i}} = 0$ | $-\mathbf{n} \cdot \mathbf{i} = 0$ | $-\mathbf{n} \cdot \mathbf{q} = 0$ | $-\mathbf{n} \cdot \mathbf{J}_{\mathrm{j}} = 0$ |

Table 5 Transport properties related to temperature and pressure.

| Parameters | Correlations ($T$ in K and $P$ in Pa) |
|---|---|
| Hydrogen dynamic viscosity (kg m$^{-1}$ s$^{-1}$) [62] | $\mu_{H_2} = 3.205\times10^{-3}\left(T/293.85\right)^{1.5}\left(T+72\right)^{-1.0}$ |
| Oxygen dynamic viscosity (kg m$^{-1}$ s$^{-1}$) [62] | $\mu_{O_2} = 8.46\times10^{-3}\left(T/292.25\right)^{1.5}\left(T+127\right)^{-1.0}$ |
| Water vapor dynamic viscosity (kg m$^{-1}$ s$^{-1}$) [62] | $\mu_{v} = 7.512\times10^{-3}\left(T/291.15\right)^{1.5}\left(T+20\right)^{-1.0}$ |
| Liquid water dynamic viscosity (kg m$^{-1}$ s$^{-1}$) [62] | $\mu_{l} = 2.414\times10^{-5}\times10^{247.8/(T-140)}$ |
| Hydrogen diffusivity (m$^{2}$ s$^{-1}$) [66] | $D_{H_2} = 1.055\times10^{-4}\left(T/333.15\right)^{1.5}\left(101325/P\right)$ |
| Oxygen diffusivity (m$^{2}$ s$^{-1}$) [66] | $D_{O_2} = 2.652\times10^{-5}\left(T/333.15\right)^{1.5}\left(101325/P\right)$ |
| Water vapor diffusivity in the anode (m$^{2}$ s$^{-1}$) [66] | $D_{v}^{a} = 1.055\times10^{-4}\left(T/333.15\right)^{1.5}\left(101325/P\right)$ |
| Water vapor diffusivity in the cathode (m$^{2}$ s$^{-1}$) [66] | $D_{v}^{c} = 2.982\times10^{-5}\left(T/333.15\right)^{1.5}\left(101325/P\right)$ |

Table 6 Design and operating parameters used in the PEMFC model.

| Parameters | Values |
|---|---|
| Anode inlet temperature (K) [55] | 353.15 |
| Cathode inlet temperature (K) [55] | 353.15 |
| Anode inlet relative humidity [55] | 84% |
| Cathode inlet relative humidity [55] | 59% |
| Stoichiometry ratio at the anode [55] | 2.0 |
| Stoichiometry ratio at the cathode [55] | 1.5 |
| Anode standard entropy change (J mol$^{-1}$ K$^{-1}$) [56] | 130.68 |
| Cathode standard entropy change (J mol$^{-1}$ K$^{-1}$) [56] | 32.55 |
| Phase transition coefficient of liquid water (s$^{-1}$) [55] | 100 |
| Operating pressure (outlet pressure) (atm) [55] | 1.0 |
| Operating voltage (V) | 0.6/0.2 |

The stoichiometric ratios used in this study are set mainly to be consistent with the working conditions in the experimental study used for the validation [67] and to ensure the adequate supply of anode hydrogen to reduce the impact of anode parameters on cell performance, highlighting the impact of cathode parameters on cell performance.

### 2.5 3D model simulation method

The conservation equations of the 3D PEMFC models are solved using the commercial software COMSOL Multiphysics 5.6 with the finite element method (FEM). All computational domains are discretized using structured grids. Since the accuracy of the simulation may be affected by the number of grid elements, a grid independence study is performed by progressively refining the mesh. The polarization curves calculated under different grid cell numbers are compared, as shown in Fig. 2a. Finally, the number of divided grids is 110000. The maximum difference in cell performance is less than 1%, which proves that the grid resolution is reasonable. The numerical model has a relative tolerance of $10^{-4}$ (or $10^{-3}$ in some cases for convergence).

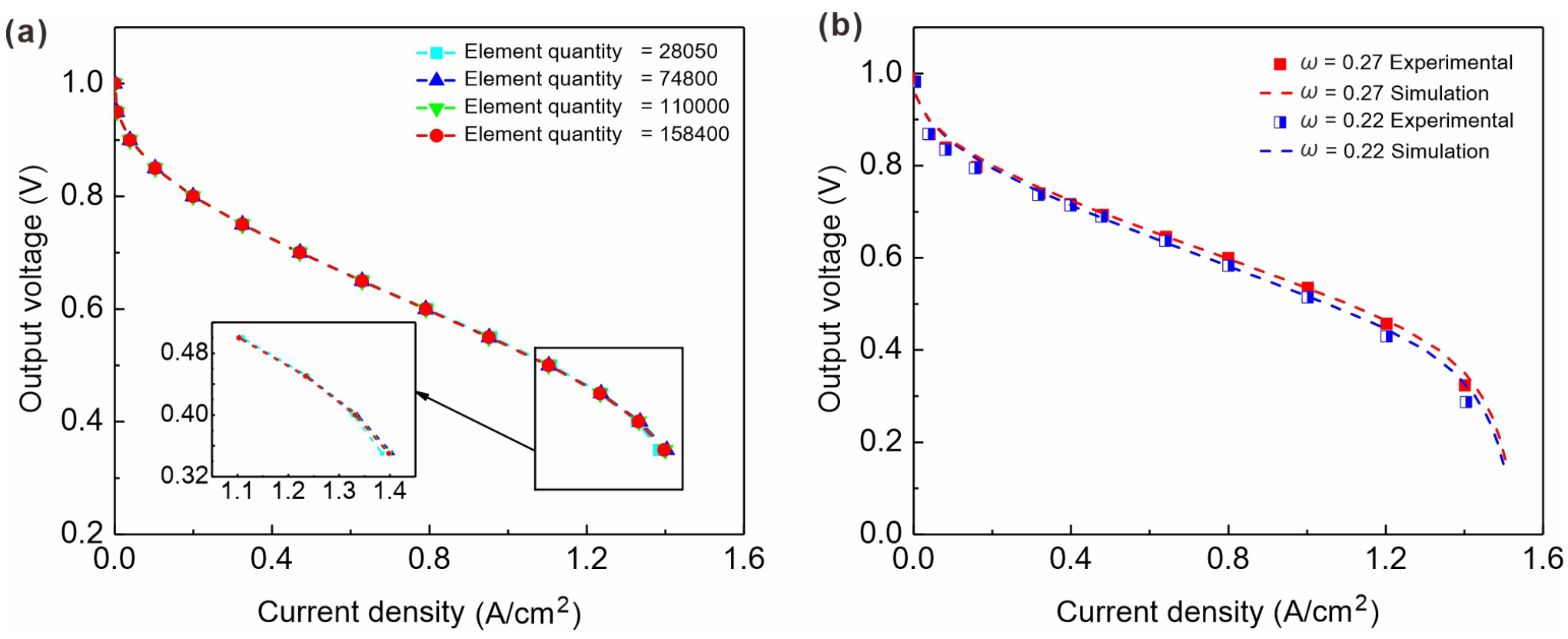

Fig. 2. (a) Grid independence study of PEMFC simulation. The ionomer volume fraction $\omega$ in the CL is 0.27. (b) Comparison between simulation results and experimental data on polarization curves under different ionomer volume fractions ($\omega$) in the CL.

### 2.6 Validation of the simulation model

To ensure the accuracy and reliability of the simulation, experimental data are used to validate the model. The simulated polarization curves under different ionomer volume fractions ($\omega = 0.22, 0.27$) in the CL are compared with the experimental data, as shown in Fig. 2b. The design parameters and operating conditions of the PEMFC numerical model are consistent with the experimental conditions in Ref. [67]. It can be found in Fig. 2b that the simulation results and the experimental results are in good agreement, which proves that the model is reliable.

## 3. Results and discussion

### 3.1 Oxygen transport and water removal in CFRFF

A comparison is carried out for the cell performance of the different cathode structures in Fig. 3a. The peak power density of the CFRFF is 3.90% higher than that of the CRFF, and the limiting current density increases by 18.09%. It is because the existence of the MFRs enhances the oxygen transfer and the water removal capability under the ribs, effectively relieving the accumulation of water under the ribs, and preventing the blockage of capillary pores caused by liquid water in porous medium (such as the CL, MPL, GDL, and MFR). As shown in Fig. 3b, the maximum pressure drop of the CFRFF is 30.64% lower than that of the CRFF, and the pumping power is smaller. The reason is that the CFRFF increases the flowing space and provides more pathways for gas flowing, thus decreasing the cathode pressure drop. Therefore, the CFRFF design is beneficial for the performance improvement of large-scale PEMFC design by enhancing under-rib mass transport and water removal capability without increasing pumping power.

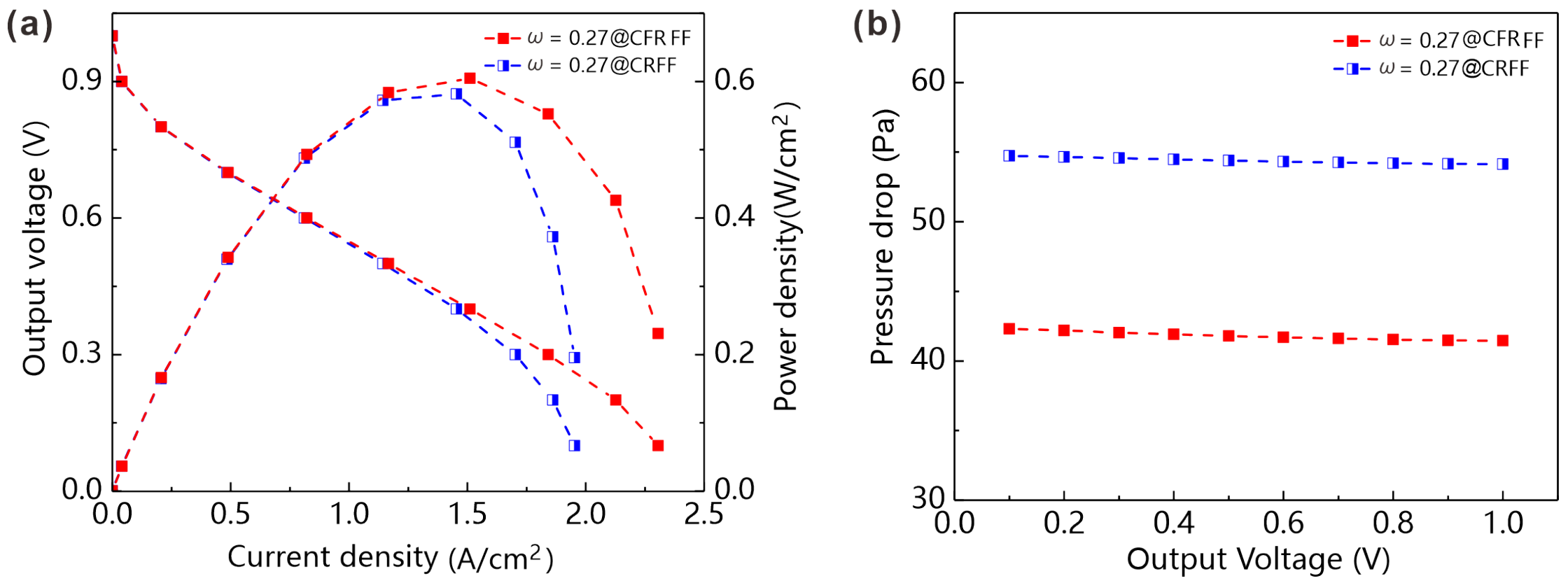

Fig. 3. Comparison of the cell performance between the CFRFF and the CRFF: (a) polarization curves and power density curves; (b) cathode pressure drop.

To further confirm that the oxygen transport capability under the rib is enhanced by the CFRFF, the oxygen concentration distribution at the cathode CL-MPL interface is compared, as shown in Fig. 4a and 4b. The oxygen concentration uniformity index ($U_{O2}$) of the CFRFF structure and the CRFF structure are 0.90 and 0.87 at 0.6 V and 0.46 and 0.31 at 0.2V, respectively, indicating that the CFRFF flow field structure has a more uniform oxygen concentration distribution. Due to the consumption by the electrochemical reaction, the oxygen concentration distribution shows an inevitable decreasing trend along the flow direction. The oxygen concentration under the ribs for the CFRFF is higher than that for the CRFF, and the distribution is more uniform. As the operating voltage decreases from 0.6 to 0.2 V, the reaction is enhanced and the oxygen consumption increases, leading to under-rib oxygen starvation in the CRFF. However, for the CFRFF, the oxygen concentration is higher and its distribution is more uniform than that for the CRFF. It is because the MFR structure of the CFRFF changes the CRFF gas flow pattern under the ribs. In the CFRFF, the reaction gas diffuses directly from the MFRs to the reaction sites in the CL under the ribs. Therefore, the gas diffusion path is shortened and the gas diffusion resistance is reduced, compared with the CRFF, in which the diffusion is from the channel to the reaction site in the CL under the ribs. The oxygen concentration along the flow direction for the CFRFF and the CRFF is compared in Fig. 4c, which indicates that a higher oxygen concentration can still be maintained in the downstream at 0.6 and 0.2 V for the CFRFF, compared with the CRFF. The average oxygen concentration at the CL-MPL interface is shown in Fig. 4d. As the operating voltage decreases, the average oxygen concentration for the CFRFF drops slower than that for the CRFF. This is because the oxygen transport resistance in the CFRFF is smaller than that in the CRFF, hence, the oxygen starvation under the ribs is relieved for the CFRFF. These results show that the CFRFF has a better mass transport capability than the CRFF, and can effectively alleviate the problem of insufficient oxygen under the ribs at large current densities.

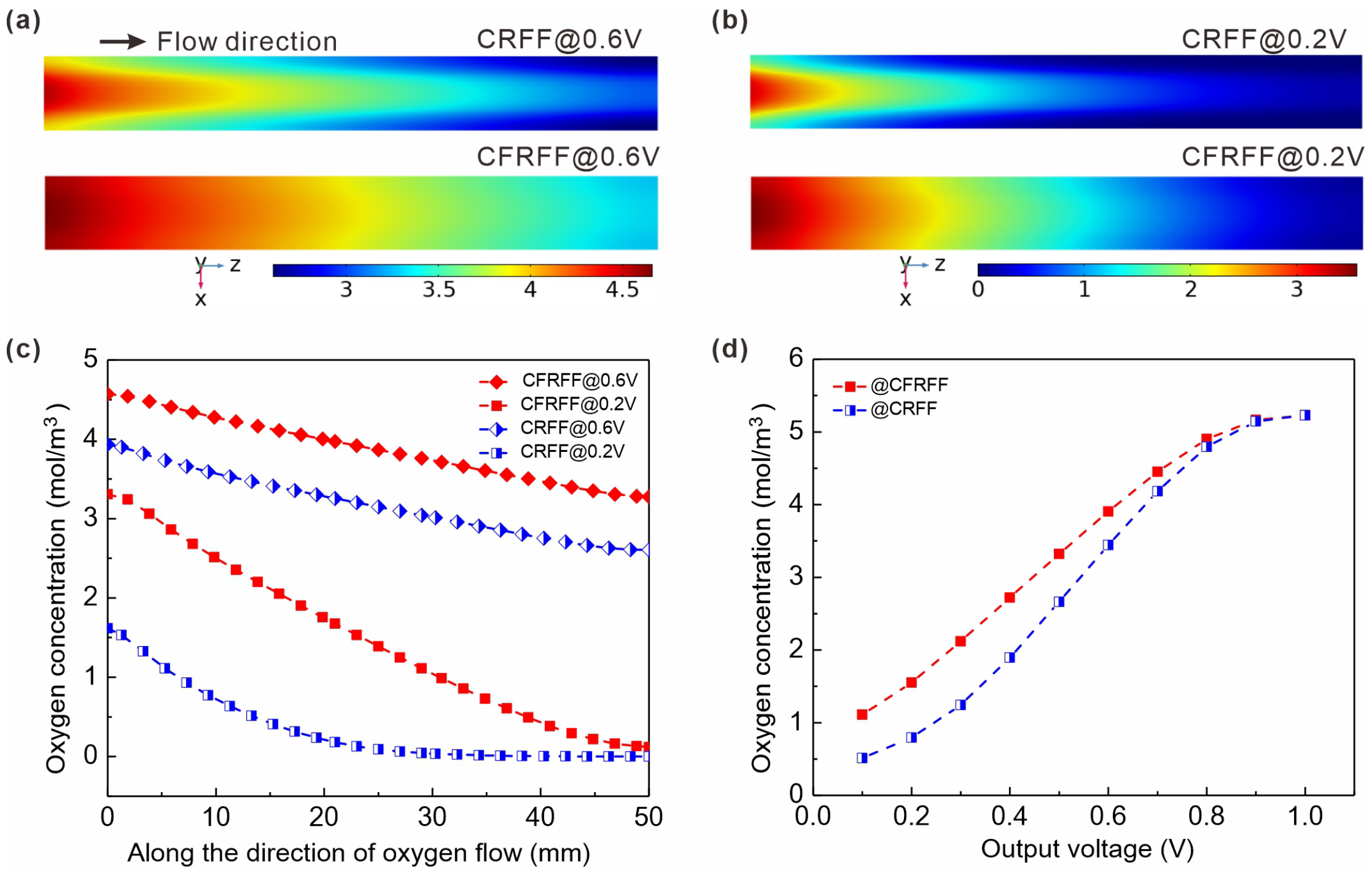

Fig. 4. Comparison of oxygen concentration distribution between CFRFF and CRFF at the cathode CL-MPL interface at (a) 0.6 V and (b) 0.2 V; (c) oxygen concentration along the flow direction ($x$ = 0.7 mm, $y$ = 1.73 mm, under the rib); (d) the average oxygen concentration at the cathode CL-MPL interface.

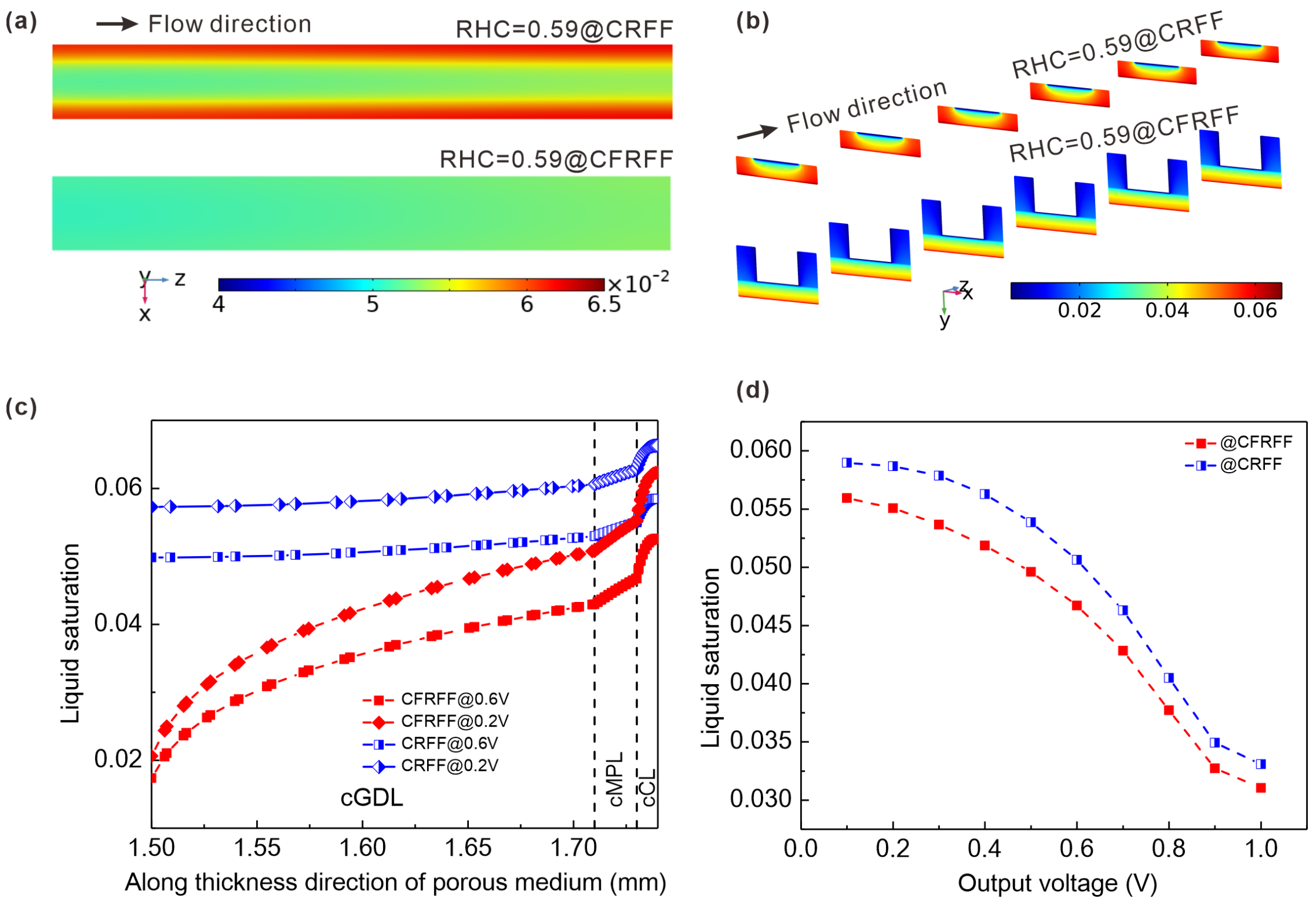

Fig. 5. (a) Comparison of the liquid water saturation distribution between the CFRFF and the CRFF at the cathode CL-MPL interface. (b) Liquid water saturation distribution at different cross sections along the flow direction of the porous medium (CL, MPL, GDL, and MFR). (c) Liquid water saturation distribution along the thickness direction of the porous medium ($x$ = 0.7 mm, $z$ =25 mm, under rib at the middle position of the flow direction). (d) Average liquid water saturation for the CFRFF and the CRFF at different operating voltages.

The CFRFF also shows better under-rib water removal performance than the CRFF. Due to the existence of liquid water in the porous medium (CL, MPL, GDL, and MFR), it may block the pores and increase the mass transport resistance of the reaction gas. The liquid water saturation distribution at the cathode CL-MPL interface is shown in Fig. 5a. For the CRFF, liquid water often accumulates under the ribs and blocks the pores, resulting in inefficient mass transfer and low oxygen concentration. For the CFRFF, the liquid water mainly concentrates under the ribs in the downstream of PEMFC. Although the distribution trend is similar to that of CRFF, both the area of water accumulation and the value of liquid water saturation are much smaller than those of CRFF. As shown in Fig. 5b and 5c, the liquid water generated by the electrochemical reaction is driven by the liquid pressure and moves to the MFR area in the CFRFF, which increases the effective drainage area of the GDL-channel interface and reduces the accumulation of liquid water under the ribs of the GDL. The average liquid water saturation at the CL-MPL interface is shown in Fig. 5d. As the operating voltage decreases, the liquid water saturation in the CFRFF increases more slowly relative to the CRFF. That is, the CFRFF has a better water-removal capability than the CRFF at different operating voltages. Therefore, the CFRFF can reduce the liquid water saturation under the ribs, strengthens the oxygen transfer capability, and improves the electrochemical reaction.

Fig. 6a and 6b show the dissolved water content distribution for the CFRFF and the CRFF at the CL-MPL interface. Due to the influence of gas-liquid transport in the flow field, the dissolved water content in the downstream of the channel is relatively high, and the dissolved water content in the under-channel region is lower than that in the under-rib region. It can be seen that the dissolved water content for the CFRFF is lower than the CRFF, and the dissolved water content distribution for the CFRFF is more uniform, no matter at 0.6 or 0.2 V. This is because the CRFF has better gas-liquid transport, which is conducive to the desorption of dissolved water. As shown in Fig. 6c, the proton transport for the CFRFF and the CRFF is enhanced with the decrease of the operating voltage, which promotes the electroosmotic transport of dissolved water from the anode to the cathode, and enhances the dissolved water content of the cathode ionomer. However, due to the better water removal capability of the CFRFF, the dissolved water content in the ionomer is reduced even more, especially in the CCL. In addition, the average dissolved water content at the CL-MPL interface is shown in Fig. 6d. As the operating voltage decreases, the dissolved water content for the CFRFF increases more slowly relative to CRFF. Considering the increasing trend of the average liquid water saturation at the CL-MPL interface at different operating voltages, we can see that the water removal capability for the CFRFF is better than that for the CRFF.

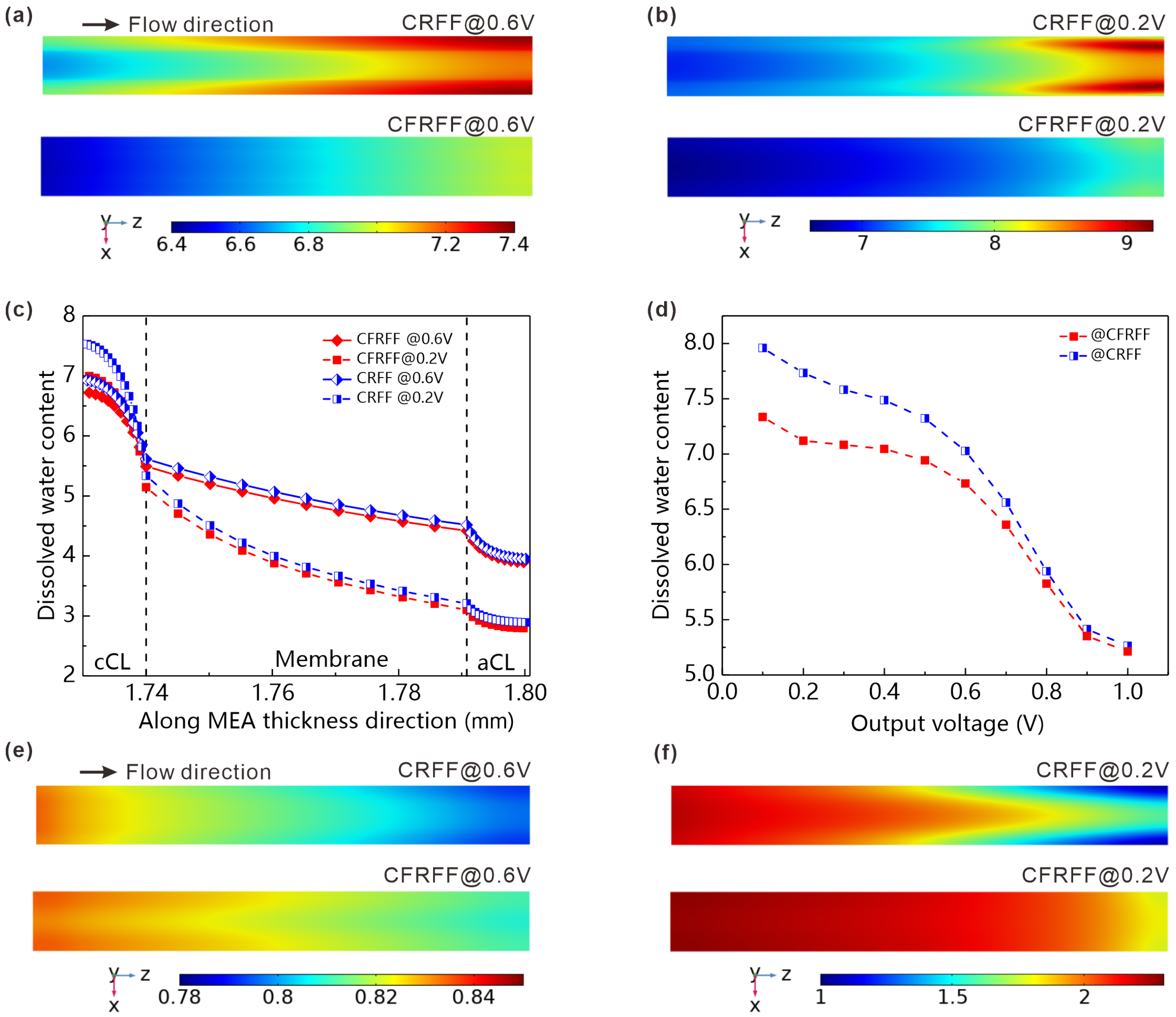

Fig. 6. (a, b) Comparison of the dissolved water content distribution between the CFRFF and the CRFF at the CL-MPL interface at 0.6 V and 0.2 V; (c) dissolved water content distribution along the MEA thickness direction ($x$ = 0 mm, $z$ = 25 mm, under the channel); (d) average dissolved water content for the CFRFF and the CRFF at the CL-MPL interface at different operating voltages; (e, f) comparison of the current density distributions between the CFRFF and the CRFF at the CL-MPL interface at 0.6 V and 0.2 V.

The current density distribution is affected directly by the oxygen concentration and potential distribution, and indirectly by the dissolved water content and liquid water saturation distribution. The current density distribution can reflect the effect of the flow field on the mass transport and charge transport of the reactants. In Fig. 6e, the local current densities of the CFRFF and the CRFF both decrease along the flow direction, and the maximum current density for the CFRFF appears in the under-rib region at the inlet. In contrast, the maximum current density for the CRFF appears in the under-channel region at the inlet because of the higher oxygen concentration under the channel. This result indicates that at 0.6 V, the effect of the potential distribution on the local current density is dominant, and the under-rib region has a strong mass transfer capability, which avoids oxygen starvation under the ribs. The local current density distributions for the CFRFF and the CRFF at 0.2 V are shown in Fig. 6f. Due to the enhanced electrochemical reaction, the local current density distributions for the CFRFF and the CRFF are mainly affected by the oxygen concentration. The local current density under the channel is higher than that under the rib, and the local current density distribution for the CFRFF is more uniform than that for the CRFF. This result further proves the superior oxygen transfer performance of the CFRFF structure.

The uniformity index of the temperature field for the CFRFF and CRFF are 0.997 and 0.997 at 0.6 V, and 0.95 and 0.91 at 0.2 V, respectively. This suggests that the temperature field distribution of CFRFF is more uniform than that of CRFF, as shown in Fig 7a and b. The temperature of the CFRFF structure is higher compared with the CRFF structure. There is a strong correlation between the temperature distribution and the electrochemical reaction intensity for the PEMFC. The CFRFF has better cell performance and a faster electrochemical reaction rate at the cathode CL-MPL interface. Therefore, the electrochemical reaction also releases more heat, thus increasing the temperature at the interface in the CFRFF. On the other hand, due to the change in gas transfer caused by the CFRFF, the oxygen transfer capability is enhanced, the oxygen distribution uniformity at the interface is improved, and the current density distribution is further improved. Thus, the temperature distribution uniformity of the fuel cell is improved, and the formation of local hot spots is avoided. This is beneficial to improve the durability of the fuel cells.

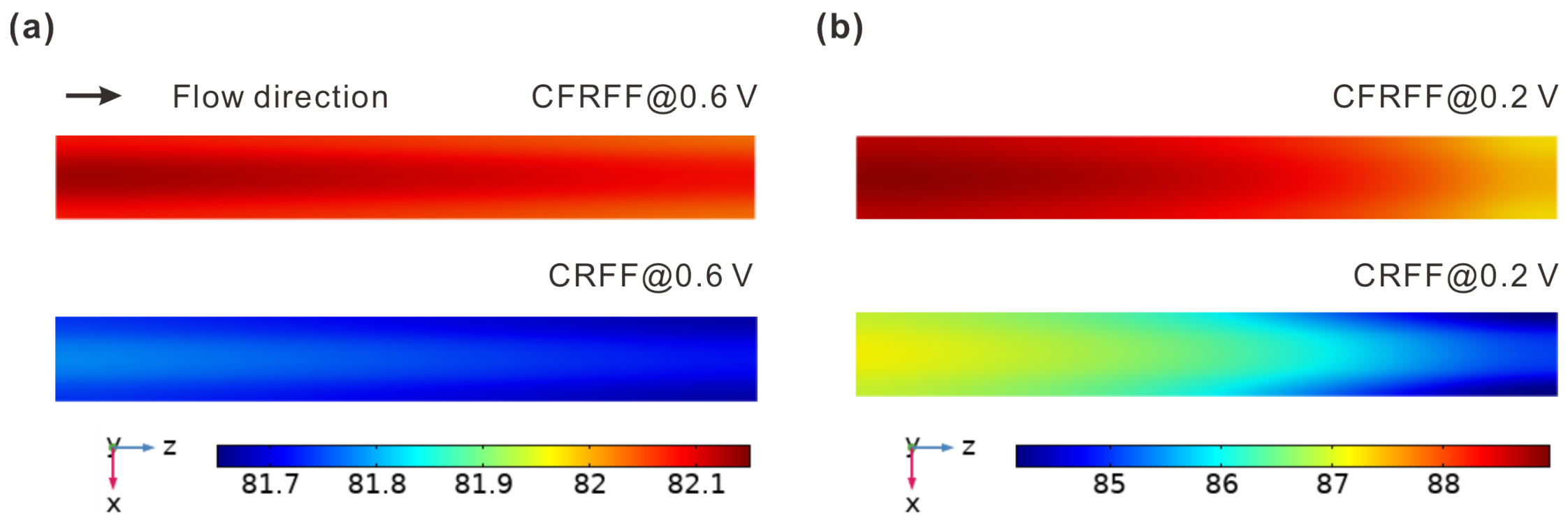

Fig. 7. Comparison of the temperature distribution between CFRFF and CRFF at the cathode CL-MPL interface at (a) 0.6 V and (b) 0.2 V.

From the above analyses, we can have a clear understanding of the mechanism for cell performance enhancement in the CFRFF. The MFRs in the CFRFF can change the under-rib oxygen transfer path, increase the effective drainage area in channels, and reduce water content inside the porous medium. Therefore, they can increase the oxygen concentration in the catalyst layer, thus enhancing the electrochemical reaction.

### 3.2 Cell performance under different operating conditions

The cell performance of CFRFF and CRFF are compared and analyzed under conventional fuel cell operating conditions and analysis range [4, 68-71]. Gas humidification is an important method to increase the dissolved water content of PEM, which affects both the electrochemical reaction and the oxygen mass transfer. The effects of inlet relative humidity in the cathode (RHC) on the cell performance for the CFRFF and the CRFF are analyzed, as shown in the polarization curves and the power density curves in Fig. 8a. With the increase of RHC, the peak power density for the CFRFF and the CRFF also increases, while the limiting current density decreases. This is because the increase in the relative humidity leads to an increase in the dissolved water content in the membrane, which in turn reduces the ohmic loss. In addition, the water content in the porous medium (CL, MPL, GDL, and MFR) also increases, which blocks the pores and increases the concentration loss.

The stoichiometric ratio in the cathode (SRC) is another important factor affecting the mass transport and water removal capability. The influence of SRC on the cell performance can be seen from the polarization curve and the power density curve in Fig. 8b. With the increase in the SRC, the peak power density and the limiting current density also increase. The higher airflow results in a higher oxygen concentration at the reaction sites in the CL, and hence increases the reaction rate. In addition, the higher airflow helps to remove excess liquid water, which alleviates the flooding phenomenon.

The operating pressure and the operating temperature are important factors for the water vapor transport in PEMFC, as shown in the polarization curve and the power density curve in Fig. 8c and 8d. With increasing the operating pressure, both the peak power density and the limiting current density increase, promoting the transfer of oxygen to the CL, thereby enhancing the electrochemical reaction intensity. With increasing the operating temperature, the peak power density and limit current density first increase and then decrease, and the optimal operating temperature is 80 °C. Although the increase in operating temperature can enhance the electrochemical reaction activity to improve cell performance, it also increases the dehydration of the membrane, reduces the dissolved water content, and increases the ion transport resistance. When the operating temperature is higher than 80 °C, the increase of ionic resistance caused by gas-liquid transport dominates, which reduces the cell performance, compared with the decrease of the reaction energy barrier.

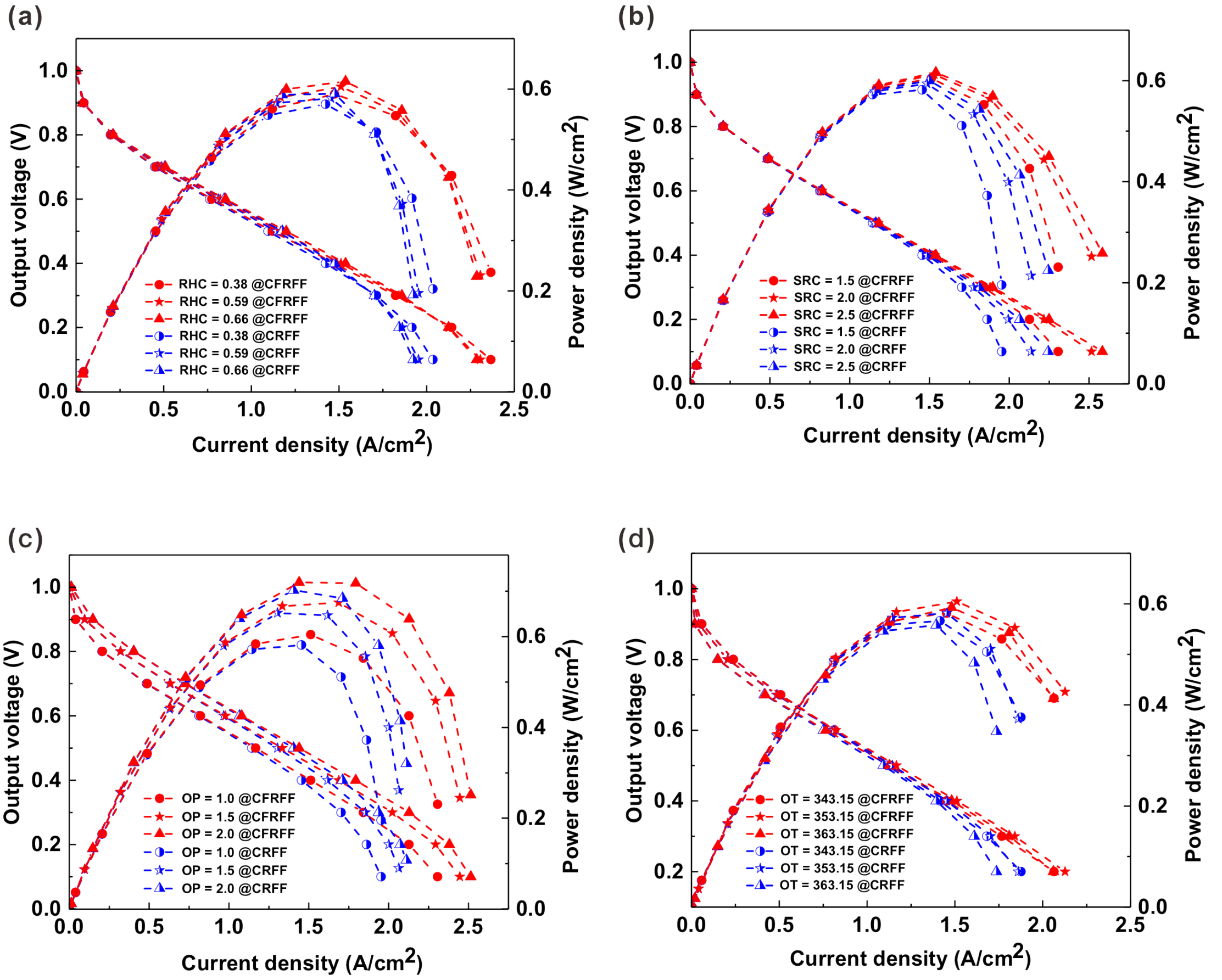

Fig. 8. Comparison of the cell performance between the CFRFF and the CRFF under various operating conditions: (a) inlet relative humidity in the cathode (RHC), (b) stoichiometric ratio in the cathode (SRC), (c) operating pressure (OP), and (d) operating temperature (OT).

By comparing the cell performance at different operating conditions, we can see that all the peak power densities and the limiting current densities for the CFRFF are higher than that for the CRFF, confirming the significant improvement of the cell performance under different operating conditions for the CFRFF compared with the CRFF.

### 3.3 Optimization of the CFRFF structural parameters

To optimize the CFRFF structure, a comparison is carried out for different metal foam filling ratios in the CFRFF with the peak power density as the objective for optimization, as shown in Fig. 9a. The metal foam filling ratio ($R$) is defined as the ratio of the metal foam rib thickness and the total rib thickness (including the metal foam rib and the solid rib), $R = H_{\mathrm{MFR}} / H_{\mathrm{rib}}$. Hence, $R$ = 0 corresponds to a CRFF, and $R$ = 1 corresponds to a flow field with only metal foam rib (without solid rib).

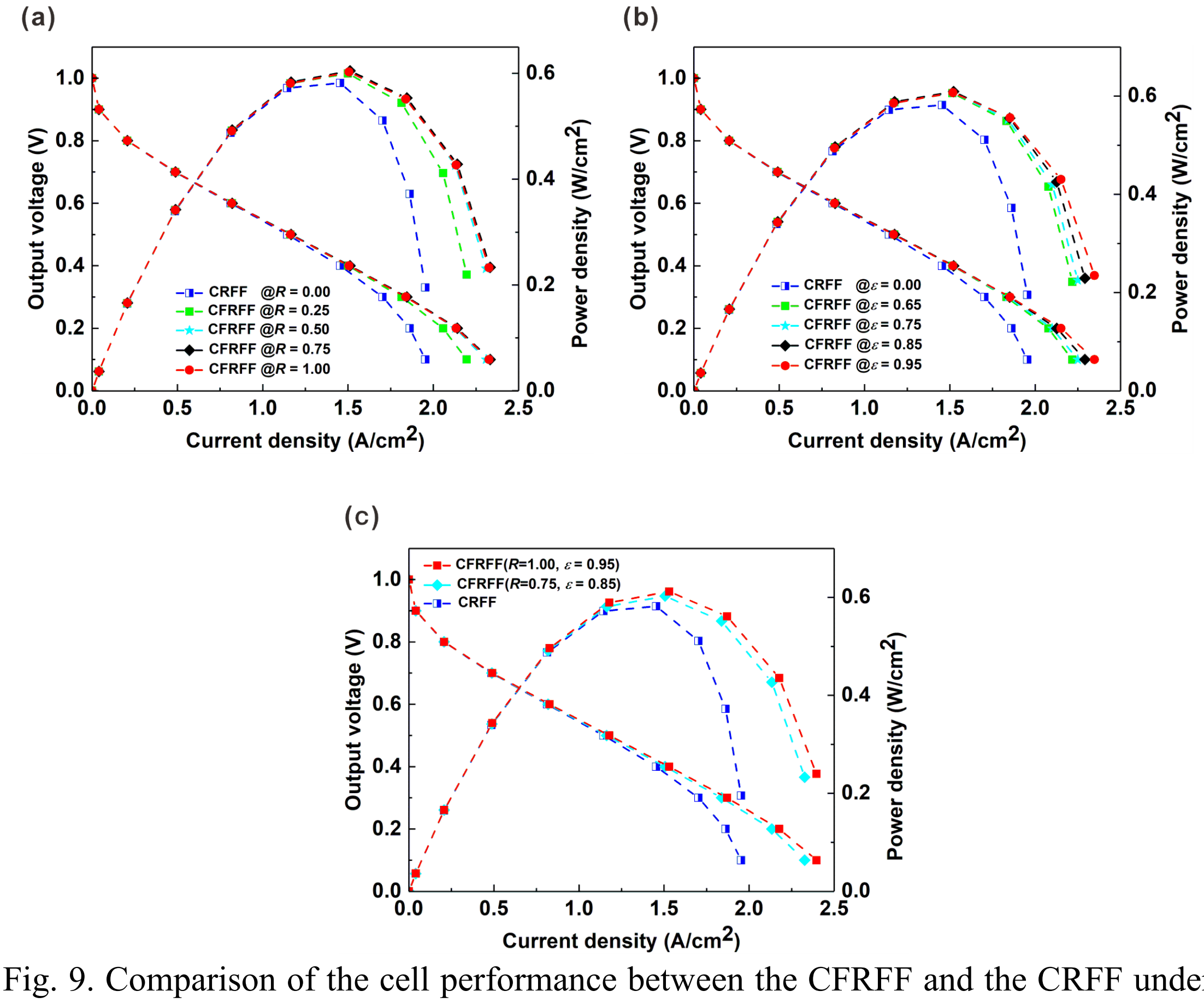

Fig. 9. Comparison of the cell performance between the CFRFF and the CRFF under different structural parameters: (a) effect of the metal foam filling ratio ($R$) with $\varepsilon$ = 0.95; (b) effect of the metal foam porosity ($\varepsilon$) with $R$=0.5. (c) Comparison between the CFRFF with the optimized structural parameters ($R$=0.75 and $\varepsilon$ = 0.85) and the CRFF.

When the filling ratio increases from 0 to 0.5, the peak power density and the limiting current density increase rapidly. That is because the composite foam-rib structure changes the mass transfer mechanism compared with the conventional flow field structure, resulting in a rapid increase in cell performance. When the filling ratio further increases from 0.5 to 1.0, the cell performance increases and then decreases slightly. The optimal filling ratio is determined to be 0.75, in which condition, the peak power density and the limiting current density of the CFRFF are 3.95% and 19.43% higher than that of the CRFF, respectively.

The effect of the metal foam porosity on the cell performance is dominated by the concentration loss, as shown in Fig. 9b. The change in the metal foam porosity can affect the electrical conductivity, the intrinsic permeability of MFRs, the gas diffusion coefficient, and therefore, the gas transport and liquid water removal capabilities under the ribs. The effect of the porosity on the permeability of the MFR is considered by using the Blake-Kozeny equation [15]. As the porosity decreases, the oxygen transport and water removal capabilities under the MFRs area are weakened, the concentration loss is increased, and the limiting current density is reduced. When the porosity is 0.95, the limiting current density of the CFRFF is the highest. However, for the peak power density, when the porosity changes from 0 to 0.85, the oxygen transfer capability for the CFRFF flow field increases, which enhances the electrochemical reaction and leads to the increase of peak power density. When the porosity further increases from 0.85 to 0.95, more water is removed in the CFRFF flow field, which reduces the dissolved water content in the membrane and increases the ion transport resistance, thus reducing the peak power density. The optimal porosity (corresponding to the highest peak power density) is determined to be 0.85. In this condition, the peak power density and the limiting current density of the CFRFF are 4.66% and 17.32% higher than that of the CRFF, respectively.

Based on the above analysis, the optimal metal foam filling ratio and porosity were combined into the model for coupling calculation. As shown in Fig. 9c, the peak power density and the limiting current density of the CFRFF are 5.20% and 22.68% higher than that of the CRFF, respectively. The peak power density and the limiting current density of the CFRFF are 2.00% and 3.00% higher than that of the CFRFF ($R$=1.00 and $\varepsilon$ = 0.95), respectively.

## 4. Conclusions

This study proposes a composite foam-rib flow field (CFRFF) structure, which combines the metal foam flow field and the conventional rib flow field. The oxygen transport and the water removal are investigated through a 3D homogeneous non-isothermal PEMFC numerical model. The results show that the CFRFF can improve cell performance by enhancing under-rib mass transport and water removal capabilities. In addition, the effects of the key operating conditions and the structural parameters are analyzed. The conclusions are summarized as follows:

(1) The CFRFF can change the gas flow pattern and increase the water removal area under the ribs through the metal foam. Thus, it can improve the oxygen transfer and water removal capabilities under the ribs, improve the oxygen concentration and current density without increasing the pumping power, and improve the cell performance.

(2) Both the peak power density and the limiting current density for the CFRFF are higher than the conventional rib flow field (CRFF) under different operating conditions, such as different relative humidity in the cathode, different stoichiometric ratios in the cathode, different operating pressure, different operating temperature. It is further confirmed that the CFRFF has better mass transport and water removal capacity and cell performance than the CRFF.

(3) The structural parameters of the CFRFF are optimized. The optimal filling ratio of metal foam in the CFRFF and the optimal porosity are 0.75 and 0.85, respectively. In this condition, the peak power density of the CFRFF is 5.20% higher than that of the CRFF, and the limiting current density is increased by 22.68%.

## Acknowledgments

This work is supported by the National Natural Science Foundation of China (Grant Nos. 51920105010 and 51921004), and the Department of Science and Technology of Inner Mongolia (Grant No. 2022JBGS0027).

## Conflict of interests

The authors declare that they have no known competing financial interests or personal relationships that could have appeared to influence the work reported in this paper.